\documentclass[aps,prl,twocolumn,showpacs,superscriptaddress]{revtex4-1}  
\usepackage{graphicx}  
\usepackage{dcolumn}   
\usepackage{bm}        
\usepackage{comment}
\usepackage{amssymb}   
\usepackage{amsmath}
\usepackage{pdfpages}
\begin{document}

\preprint{APS/123-QED}

  \title{Binary Black Hole Mergers from Globular Clusters: Implications for Advanced LIGO}
  \author{Carl L.\ Rodriguez}  
  \affiliation{Center for Interdisciplinary Exploration and Research in
    Astrophysics (CIERA) and Dept.~of Physics and Astronomy, Northwestern
      University, 2145 Sheridan Rd, Evanston, IL 60208, USA}
  \author{Meagan Morscher}
    \affiliation{Center for Interdisciplinary Exploration and Research in
    Astrophysics (CIERA) and Dept.~of Physics and Astronomy, Northwestern
      University, 2145 Sheridan Rd, Evanston, IL 60208, USA}
  \author{Bharath Pattabiraman}
    \affiliation{Center for Interdisciplinary Exploration and Research in
    Astrophysics (CIERA) and Dept.~of Physics and Astronomy, Northwestern
      University, 2145 Sheridan Rd, Evanston, IL 60208, USA}
      \affiliation{Dept.~of Electrical Engineering and Computer Science, 
      Northwestern University, Evanston, IL, USA}
  \author{Sourav Chatterjee}
    \affiliation{Center for Interdisciplinary Exploration and Research in
    Astrophysics (CIERA) and Dept.~of Physics and Astronomy, Northwestern
      University, 2145 Sheridan Rd, Evanston, IL 60208, USA}

      \author{Carl-Johan Haster}
         \affiliation{Center for Interdisciplinary Exploration and Research in
    Astrophysics (CIERA) and Dept.~of Physics and Astronomy, Northwestern
      University, 2145 Sheridan Rd, Evanston, IL 60208, USA}
     \affiliation{School of Physics and Astronomy, University of Birmingham, Birmingham, B15 2TT, United Kingdom}
     
  \author{Frederic A.\ Rasio}
    \affiliation{Center for Interdisciplinary Exploration and Research in
    Astrophysics (CIERA) and Dept.~of Physics and Astronomy, Northwestern
      University, 2145 Sheridan Rd, Evanston, IL 60208, USA}

\date{\today}

\begin{abstract}
The predicted rate of binary black hole mergers from galactic fields can vary over several orders of magnitude and is extremely sensitive to the assumptions of stellar evolution.   But in dense stellar environments such as globular clusters, binary black holes form by well-understood gravitational interactions.  In this letter, we study the formation of black hole binaries in an extensive collection of realistic globular cluster models.    By comparing these models to observed Milky Way and extragalactic globular clusters, we find that the mergers of dynamically-formed binaries could be detected at a rate of $\sim 100$ per year, potentially dominating the binary black hole merger rate.  We also find that a majority of cluster-formed binaries are more massive than their field-formed counterparts, suggesting that Advanced LIGO could identify certain binaries as originating from dense stellar environments.
\end{abstract}

\maketitle


\section{Introduction}
\label{sec:intro}

By the end of this decade, the Advanced LIGO and Virgo detectors are expected to
observe gravitational waves (GWs), ushering in a new post-electromagnetic era of
astrophysics \cite{Aasi2015,Acernese2015}.  The most anticipated sources of
observable GWs  will be the signals generated by mergers of binaries with
compact object components, such as binary neutron stars (NSs) or binary black
holes (BHs).  While coalescence rates of NS-NS or BH-NS systems can be
constrained from observations, it is not currently possible to produce
observationally-motivated rate predictions for BH-BH mergers
\cite{Belczynski2002}.  Typical detection rates of binary BH (BBH) mergers in
galaxies can span several orders of magnitude from $0.4\ \rm{yr}^{-1}$ to $1000\
\rm{yr}^{-1}$ with a fiducial value of $\sim 20\ \rm{yr}^{-1}$ \cite{Abadie2010}; however, these estimates typically ignore the large numbers of BBHs that are formed through dynamical interactions in dense star clusters  \cite{Morscher2015,Sadowski2007a}.

The dynamical formation of BBHs is a probabilistic process, requiring a very high stellar density.  These conditions are believed to exist within the cores of globular clusters (GCs), very old systems of $\sim 10^5-10^6$ stars with radii of a few parsecs.  Approximately 10 Myr after the formation of a GC, the most massive stars explode as supernovae, forming a population of single and binary BHs with individual masses from $\sim 5 M_{\odot}$ to $\sim 25M_{\odot}$ \cite{Belczynski2006}.  The BHs, being more massive than the average star in the cluster, sink to the center of the GC via dynamical friction, until the majority of the BHs reside in the cluster core \cite{Spitzer1969}.  After this ``mass segregation'' is complete, the core becomes sufficiently dense that three-body encounters can frequently occur \cite{Ivanova2005},  producing BBHs at high rates.  In effect, GCs are dynamical factories for BBHs: producing large numbers of binaries within their cores and ejecting them via energetic dynamical encounters.

In this letter, we use an extensive and diverse collection of GC models to study
the population of BBHs that Advanced LIGO can detect from GCs.  We explore how
the observed parameters of a present-day GC correlate with the distribution of
BBH inspirals it has produced over its lifetime.  We then compare our models to
the observed population of Milky Way GCs (MWGCs) and use recent measurements of
the GC luminosity function to determine a mean number of BBH inspirals per GC.
Finally, we combine these estimates with an updated estimate of the spatial
density of GCs in the local universe (Appendix I) into a double integral over comoving volume and inspiral masses to compute the expected Advanced LIGO detection rate.  We assume cosmological parameters of $\Omega_M = 0.309$, $\Omega_\Lambda = 0.691$, and $h = 0.677$, consistent with the latest combined Planck results \cite{PlanckCollaboration2015}.

\section{Computing the Rate}
\label{sec:meth}

We use a collection of 48 GC models generated by our Cluster Monte Carlo (CMC)
code, an orbit-averaged H{\'{e}}non-type Monte Carlo code for collisional
stellar dynamics \cite{Joshi1999}.  The models span a range of initial star
numbers ($2\times 10^5$ to $1.6\times10^6$), initial virial radii (0.5 pc to 4
pc), and consider low stellar metallicities ($Z = 0.0005,0.0001$) and high
stellar metallicities ($Z = 0.005$).  In addition, the code implements dynamical
binary formation via three-body encounters, strong three and four-body binary
interactions, and realistic single and binary stellar evolution.  See
Appendix II for a complete description of our code and the models used.

Previous studies have explored the contribution of BBHs from GCs to the Advaned LIGO detection rate \cite{Zwart1999,OLeary2006,Banerjee2010,Downing2011,Bae2014,Tanikawa2013}; however, the majority of these studies have relied on either approximate analytic arguments or simplified $N$-body models with $N\lesssim 10^5$ particles and have assumed a single black hole mass of $10 M_{\odot}$.  The one exception is \cite{Downing2011}, which used a Monte Carlo approach to model GCs of a realistic size ($N=5\times 10^5$).  However, their study only considered GCs of a single mass, and did not extrapolate that result to the observed distribution of GCs in the local universe.  Ours is the first study to compare models with all the relevant physics over a range of masses to observed GCs.  This comparison enhances our BBH merger rate by more than an order of magnitude over previous results.

We express the rate of detectable mergers per year, $R_d$, as the following double integral over binary chirp mass ($\mathcal{M}_c \equiv (m_1 m_2)^{3/5} / (m_1 + m_2)^{1/5}$) and redshift:
\begin{equation}
R_d = \iint \mathcal{R}(\mathcal{M}_c,z) f_d(\mathcal{M}_c,z) \frac{dV_c}{dz} \frac{dt_s}{dt_0} d\mathcal{M}_c dz~.
\label{eqn:rate}
\end{equation}

\noindent This equation is similar to that found in \cite{Belczynski2012,Belczynski2014}.  The components of Eqn.\ \ref{eqn:rate} are as follows:

\begin{itemize}
\item $\mathcal{R}(\mathcal{M}_c,z)$ is the rate of merging BBHs from GCs with chirp mass $\mathcal{M}_c$ at redshift $z$.  
\item $f_d(\mathcal{M}_c,z)$ is the fraction of sources with chirp mass $\mathcal{M}_c$ at redshift $z$ that are detectable by a single Advanced LIGO detector.
\item $dV_c/dz$ is the comoving volume at a given redshift \cite{Hogg1999}.
\item $dt_s/dt_0 = 1/(1+z)$ is the time dilation between a clock measuring the merger rate at the source versus a clock on Earth.
\end{itemize}

This letter focuses on estimating the rate, $\mathcal{R}(\mathcal{M}_c,z)$,
using our collection of GC models.  We assume the rate can be expressed as the
product of the mean number of inspirals per GC, the distribution of those
sources in $\mathcal{M}_c-z$ space, and the density of GCs in the local
universe, i.e.\  $\mathcal{R}(\mathcal{M}_c,z) = \left<N\right> \times
P(\mathcal{M}_c,z)\times \rho_{GC}$.  The spatial density of GCs in the local
universe is taken to be $\rho_{GC}=0.77\ \rm{Mpc}^{-3}$, based on recent
measurements of extragalactic GC systems \cite{Harris2013} and modern
near-infrared Schechter functions \cite{Kelvin2014}.  Note that this estimate,
computed in Appendix I, is
substantially lower than the previous estimate of $\rho_{GC}=8.4\
h^3\rm{Mpc}^{-3}$ from \cite{Zwart1999} that has been used in previous studies.  We now estimate the values of $\left<N\right>$ and $P(\mathcal{M}_c,z)$.

\section{mean number of mergers per cluster}
\label{sec:inspPerGC}

To determine the mean number of BBH inspirals produced by a GC, we use the
collection of models to explore how the present-day observable parameters of
GCs relate to the number of BBHs it has produced over its lifetime.  To quantify
the realism of a particlar model, we compare the total masses and concentrations of
our models to GCs observed in the Milky Way.  The concentrations are measured
by considering the ratio of a cluster's core radius to its half-light radius,
$R_c/R_h$.  This mass-concentration space is similar to the ``fundamental
plane'' of GCs described in \cite{McLaughlin2000}, with $R_c/R_h$ in place of
the King concentration \footnote{The King concentration is defined as the
logarithm of the core radius over the tidal radius.  We use the simpler
$R_c/R_h$, as the tidal radius can be difficult to determine observationally.}. 

\begin{figure}[]
\centering
\includegraphics[scale=0.6, trim=0in 0.05in 0in 0.05in, clip=true ]{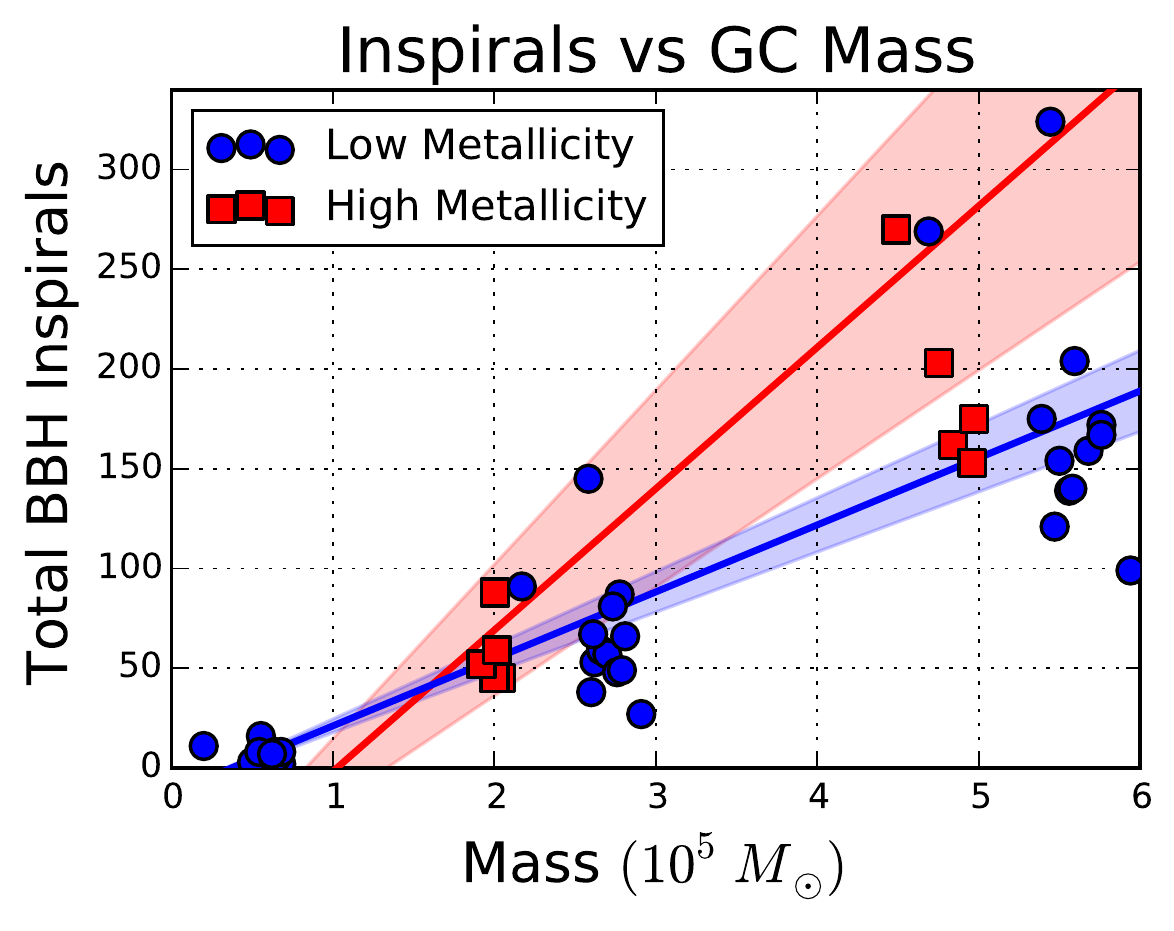}
\caption{The number of BBH inspirals per model at 12 Gyr for each of our 48 GC models as a function of final cluster mass.  We show the weighted linear regression (with $1\sigma$ uncertainties on the slope) for the low and high metallicity models.}
\label{fig:regress}
\end{figure}

Two trends emerge in our models.  First, the total number of BBH inspirals over
12 Gyr is nearly linearly proportional to the final cluster mass.  Second, the
number of inspirals is higher for more compact clusters (those with smaller
$R_c/R_h$).  Since the model coverage of the $R_c/R_h$ space is poorer than the
coverage of the mass, and since there are no detailed observations of
extragalactic GC concentrations, we elect to focus on the linear relationship
between a GC's mass and the number of inspirals it has produced.  We perform a
weighted linear regression for both low-metallicity and high-metallicity
GCs (Fig.\ \ref{fig:regress}).  The weights are created by generating a kernel
density estimate (KDE) \footnote{We select the bandwidths using maximum
likelihood cross-validation, since for multi-dimensional distributions, the
optimal bandwidth is not necessarily equal along each dimension, particularly
when using mixed units.  This is implemented in the  \protect \href
{''http://statsmodels.sourceforge.net/''}{\texttt{StatsModels}} Python package.} of the MWGCs in the fundamental plane, then measuring the probability of each model as reported by the KDE.  In other words, GC models that are more likely to represent draws from the distribution of MWGCs are more heavily weighted.

We compute the mean number of inspirals per GC by multiplying the linear relationships from Fig.\ \ref{fig:regress} by the mass distribution of GCs.  Recent work \cite{Harris2014} has suggested that the distribution of GC luminosities is universal and well-described by a log-normal distribution:

\begin{equation}
\frac{dN}{d\log L} = N_0 \exp\left(-\frac{(\log L - \log L_0)^2}{2\sigma_L^2}\right)
\label{eqn:gclf}
\end{equation}

\noindent with $\log L_0 = 5.24$ and $\sigma_L = 0.52$.  Assuming a mass-to-light ratio of 2 in solar units \cite{Harris2014,Bell2003}, we convert this luminosity function to a mass function.  We then compute the average number of inspirals per GC by integrating the linear relationship over the normalized GC mass function.  The results for both metallicities and different high-mass cutoffs are shown in Table \ref{tab:ninsp}.

\begin{table}[h]
\begin{tabular}{l|ccc}
\hline\hline
&& Mass Cutoff &\\\hline
\textbf{Metallicity}&$4\times10^6M_{\odot}$ & $2\times10^7M_{\odot}$ & $2\times10^8M_{\odot}$ \\ \hline
Low                  & 430 & 967 & 1512                           \\
High                 & 830                            & 1954 & 3103         \\ \hline\hline
\end{tabular}
\caption{The mean number of inspirals per GC over 12 Gyr.  The result depends on our choice of maximum GC mass.  We consider cutoffs of $4\times10^6M_{\odot}$ (the approximate mass of the most massive MWGC, $\omega$ Cen),  $2\times10^7M_{\odot}$ (the approximate cutoff used in \cite{Harris2014}), and $2\times10^8M_{\odot}$ (the mass of the ultra-compact dwarf M60-UCD1 \cite{Strader2013}).}
\label{tab:ninsp}
\end{table}

\section{Distribution of Inspirals}
\label{sec:inspDist}

\begin{figure}[bth!]
\centering
 \includegraphics[scale=0.6, trim=0in 0.05in 0in 0.05in,
 clip=true]{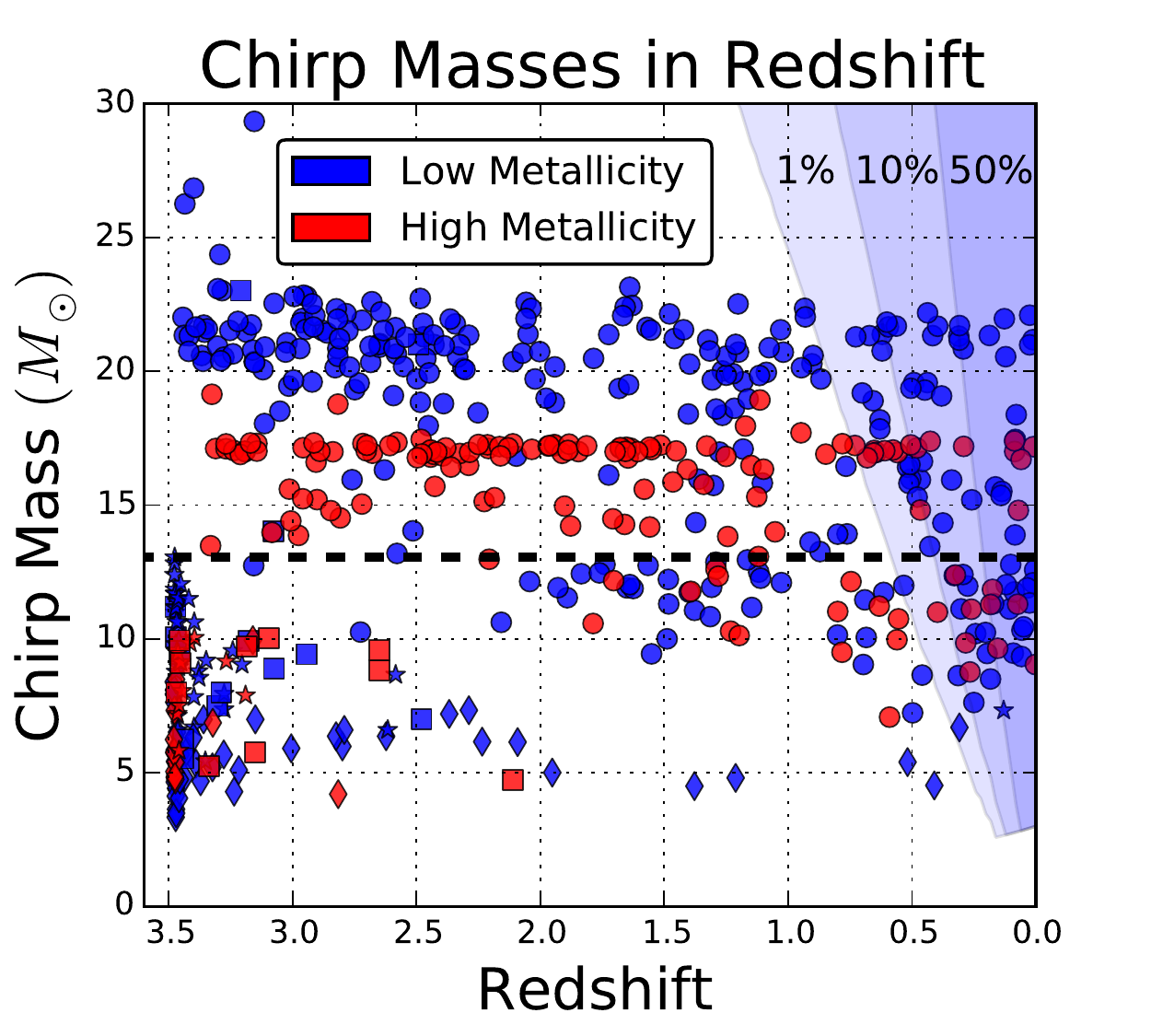}
\caption{A sampe distribution of inspirals in redshift from the set of models.
The redshift is computed by assuming that the difference between the present day
and the inspiral time corresponds to the cosmological lookback time at a given
redshift (e.g.\ \cite{Hogg1999}).  The number of
inspirals drawn from each model is proportional to its weight, or how similar it
is to the observed distribution of MWGCs.  Inspirals of BBHs that were formed
primordially are indicated with stars (merged in the cluster) and diamonds
(ejected before merger).  Inspirals of BBHs formed dynamically are shown as
squares (in-cluster) and circles (ejected).  Note that there are no binaries
that are formed by binary stellar evolution with chirp masses greater than $\sim
13M_{\odot}$ (dashed line).  This result is consistent across all models.
The
blue shaded regions illustrate the regions of parameter space where 50\%, 10\%,
and 1\% of sources are detectible by Advanced LIGO.}
\label{fig:allIns}
\end{figure}

The numbers quoted in Table \ref{tab:ninsp} provide us with the mean number of
BBH inspirals from a GC over 12 Gyr.  We could use this average rate to
compute a detection rate for Advanced LIGO.  However, it is qualitatively obvious that the mass distribution of BBH sources is not constant in time (Fig.\ \ref{fig:allIns}).  

Therefore, we must use the distribution of BBH inspiral events over time from GCs to compute the rate.  We select inspirals randomly from each of our models, drawing more inspirals from models with higher weights according to the following scheme:

\begin{equation}
W(M,R_c/R_h) = \frac{KDE_{\text{MWGC}}(M,R_c/R_h)}{KDE_{\text{Models}}(M,R_c/R_h)}
\label{eqn:weights}
\end{equation}

\noindent where the weight, $W(M,R_c/R_h)$, of a model with mass $M$ and
compactness $R_c/R_h$ at 12 Gyr is defined by the ratio of the MWGC KDE at $M,
R_c/R_h$ divided by the KDE of the models themselves, evaluated at $M,R_c/R_h$.
The reason for these weights is as follows: we wish to draw more samples from
models that are more likely to represent MWGCs, but because our collection of
models is drawn from a different distribution (the initial conditions from
\cite{Morscher2015}), we cannot simply draw inspirals at random from each model
according to how well it represents real GCs.  To do so would bias our samples
with the distribution that results from our initial conditions.  By dividing the
probability of a model representing a MWGC by the probability density of our
collection of models, our scheme naturally corrects for this. Models unlikely to
represent MWGCs have small numerators and low weights.  Models with no
neighboring models  that are likely to represent MWGCs have large numerators and
small denominators, yielding high weights.  Conversely, models \emph{with}
neighbors that are likely to represent MWGCs will have large numerators and
large denominators, yielding smaller weights; however, as we will select some
number of inspirals from each of these neighboring models, the cumulative effect is the same.

We show a sample distribution of the chirp masses versus redshift in Fig.\
\ref{fig:allIns}.  We distinguish between two different BBH formation channels:
primordial and dynamical.  We define primordial BBHs as those that are formed
from the supernovae of two main sequence stars in a binary, and whose components
were never bound to any other star before merger; conversely, we define
dynamical binaries as those that are either formed from two isolated BHs via a
three-body encounter, or formed from a higher-order dynamical encounter (a
binary-single or binary-binary interaction forming a new binary pair).
Primordial binaries can still have their orbital parameters modified by dynamics
(via a strong encounter with another BH or BBH), as long as the encounter leaves
the primordial BBH intact.  One immediately apparent feature is the bi-modality
between primordial and dynamical BBHs.  Over all of our models, the highest
chirp mass that is formed by pure binary stellar evolution is $\mathcal{M}_c\sim
13M_{\odot}$, as systems with larger progenitors are distrupted by the
supernova kick.  This implies that any source from our models with a detected chirp mass greater than $\sim 13M_{\odot}$ could \emph{only} have formed \emph{dynamically}.

To compare this result to BBHs formed in the field, we generated two additional CMC models, each containing $5\times10^6$ binaries and different metallicites ($Z = 0.005$ and $Z = 0.0005$). These models were computed without two-body relaxation, binary formation, or strong encounters, and only considered the physics of binary stellar evolution.   In this dynamics-free environment, the maximum chirp mass of any BBH was $\mathcal{M}_c \sim 13M_{\odot}$.  Although this result depends on the metallicity and the physics of stellar evolution, it does suggest that GC dynamics forms BBHs consistently more massive than those in the field

\section{Detection Rate}
\label{sec:rate}

We now compute the
expected rate of signals detectable by Advanced LIGO.  To compute the fraction
of detectable sources, $f_d(\mathcal{M}_c,z)$, we use gravitational waveforms
that cover the inspiral, merger, and ringdown phases of a compact binary merger
(known as IMRPhenomC waveforms \cite{Santamaria2010}) and compute the
signal-to-noise ratio (SNR) using the projected zero-detuning, high-power
configuration of Advanced LIGO \footnote{Both the noise curve and technical
reports describing it can be found under \protect \href
{''https://dcc.ligo.org/cgi-bin/DocDB/ShowDocument?docid=2974''}{LIGO Document
T0900288-v3}}.  We then marginalize over binary orientation and sky location to
determine what fraction of sources at a given chirp mass and redshift yield an
SNR $> 8$.  This approach is identical to that found in \cite{Belczynski2014}.
Note that we have assumed all binary components have equal masses in order to
simplify the integral.  This assumption is well-justified, as the
dynamically-formed BBHs in our models have similar component masses.  We assume all BHs to be non-spinning. 

Our distribution of inspirals, $P(\mathcal{M}_c,z)$, is generated by creating a
KDE of the inspirals in Fig.\ \ref{fig:allIns}.  
Since each ``draw'' of inspirals will produce a slightly different distribution,
we compute $P(\mathcal{M}_c,z)$ 1000 times, and then take the mean of Eqn.\
\ref{eqn:rate} for those 1000 draws.  We find that a single Advanced LIGO
detector operating at design sensitivity will detect $\sim 100$ BBHs per year
from GCs.  Of those about $2/3$ will originate from low-metallicity GCs and the
rest from high-metallicity GCs, assuming 76\% of GCs are low metallicity
(consistent with the MWGC distribution).  Approximately 80\% of these sources
will have chirp masses greater than $13(1+z)M_{\odot}$, meaning that the
majority of BBHs detectable by Advanced LIGO from GCs could only havie formed dynamically.   

{The majority of these BBH sources will be detected at low redshifts.  For
low-metallicity clusters, the distribution of detectible sources in redshift peaks at
$z\sim 0.3$, while for high-metallicity clusters the distribution peaks at
$z\sim0.24$.  In both cases, 90\% of detectable sources are located at
$z\lesssim0.57$.

To obtain a rough estimate of the uncertainty on this prediction, we perform a
simple error analysis that considers the optimistic and conservative rates that
would be obtained by varying our assumptions and selecting the $\pm 1\sigma$
estimates of certain quantities.  For the conservative estimate, we assume that
the GC mass function truncates at the mass of $\omega \text{ Cen}$
($4\times10^6 M_{\odot}$), and that the spatial density of GCs is $\rho_{GC} =
0.32\ \rm{Mpc}^{-3}$ (the conservative estimate from Appendix I).  We also recompute the rate using the $-1\sigma$ uncertainty from
the regression in Fig.\ \ref{fig:regress} and the lower standard deviation of
our 1000 draws of $\mathcal{R}(\mathcal{M}_c,z)$.  This yields a conservative
estimate of $\sim 20$ BBH inspirals per year.  Conversely, if we assume the most
optimistic truncation mass for GC mass function ($2\times10^8 M_{\odot}$), the
most optimistic GC spatial density ($\rho_{GC} = 2.3\ \rm{Mpc}^{-3}$, the
optimistic estimate from Appendix I, similar to the value used in previous studies), and the $+1\sigma$ uncertainties on the linear regression and $\mathcal{R}(\mathcal{M}_c,z)$, we find an optimistic rate of $\sim 700$ BBH inspirals per year.  This range is primarily influenced by the uncertainty in the GC spatial density and the truncation mass of the GC mass function.

\section{Conclusion}
\label{sec:conc}

In this letter, we compared new GC models computed with our CMC code to the observed distributions of Milky Way and extragalactic GCs to predict the expected rate of BBH inspirals from realistic GCs.  We determined a linear relationship between the present-day mass of a GC and the number of BBH inspirals produced by that cluster.  By combining this with the universal GC luminosity function and a new estimate for the spatial density of GCs, we were able to predict the mean density of BBH inspirals from GCs in the local universe.  Then by weighting our models according to their similarity to the observed distribution of MWGCs, we created a distribution of inspiral sources in chirp mass and redshift.  Finally, by combining this with the anticipated sensitivity of Advanced LIGO, we estimated a detection rate of $\sim 100$ BBH inspiral events per year from GC sources.  With highly conservative assumptions, this rate drops to $\sim 20$ events per year, while highly optimistic assumptions pushes the rate as high as $\sim 700$ events per year.

We also found that no BBHs with chirp masses above $\sim 13M_{\odot}$ were formed from a primordial binary.  In other words, every inspiral with $\mathcal{M}_c > 13M_{\odot}$ in our models was formed by dynamical processes alone.  This could, in theory, provide an easy way to discriminate between binaries that were formed dynamically versus those formed by binary stellar evolution; however, this result is highly dependent on the physics of supernova kicks and the fraction of ejected supernova material which falls back onto the newly formed BH, both of which remain poorly constrained.  In addition, recent work has suggested that the mass distribution of chirp masses for BBHs produced by stellar evolution can reach as high as $\mathcal{M}_c \sim 30M_{\odot}$, depending on the physics of the common envelope \cite{Belczynski2010}.  As such, this result should be treated as a proof-of-principle, and not a concrete physical claim.  Investigations to better understand the relationship between this formation cutoff, the distribution of supernovae kicks, and the fallback fraction, are currently underway.  

Finally, the number of BHs formed is entirely dependent on the choice of the initial mass function.  Although our choice of IMF is typical for studies of this type, a variation of $1\sigma$ in the slope of the high-mass end of the IMF can produce significant differences in the number of BBHs.  Investigations to quantify this effect are also underway.

\begin{acknowledgments}
We thank Ilya Mandel for his assistance in computing the fraction of detectable sources, as well as useful discussions.  We also thank Ben Farr, Claude-Andre Faucher-Giguere, William Harris, Kyle Willett, Michael Schmitt, and Tyson Littenberg for useful discussions.  CR was supported by an NSF GRFP Fellowship, award DGE-0824162.  This work was supported by NSF Grant AST-1312945 and NASA Grant NNX14AP92G.  CJH acknowledges support from an RAS grant.
\end{acknowledgments}

\bibliographystyle{apsrev}
\bibliography{main,mainNotes}{}

\appendix

\section{Number of Globular Clusters per Mpc$^3$}
\label{apx:rhogc}

In order to estimate the rate of inspirals from an average GC per $\rm{Mpc}^3$, we must compute the average spatial density of GCs in the local universe.  This can be accomplished by considering the mean number of GCs per galaxy at a given luminosity, multiplied by the spatial density of galaxies at that luminosity, then summing over all luminosities, as illustrated in the following equation:

\begin{equation}
\rho_{GC}  = \int \left(\frac{\text{\# of GCs}}{\text{Galaxy}/ \text{Mag}} \right)\times \left(\frac{\text{\# of Galaxies}}{\text{Mpc}^3 \times \text{Mag}} \right) d\text{Mag}.
\label{eqn:ngc}
\end{equation}

The number of GCs per galaxy per luminosity can be determined by use of the Harris Globular Cluster System catalog \cite{Harris2013}.  The catalog provides a list of 422 galaxies, their morphological type, visual and K-band magnitudes (where available), and the estimated total number of GCs.  In Figure \ref{fig:ngc}, we plot the 346 galaxies for which K-band photometry is available in the catalog against the estimated number of GCs.  For each collection of galaxy morphologies, we perform a Gaussian Process regression with the \texttt{George} package, described in \cite{Ambikasaran2014}.  The Gaussian processes are generated using a squared-exponential kernel combined with a white noise kernel, and then fit to the log of the number of GCs per galaxy. The kernel hyperparameters are selected by maximizing the marginalized log-likelihood of the Gaussian Process.  See \cite{Rasmussen2006} for a detailed description of regression with Gaussian Processes.  The mean and standard deviation of the resulting Gaussian Processes are also shown in Fig.\ \ref{fig:ngc}.  Note that the catalog does not include low-luminosity dwarf and irregular galaxies for which $N_{GC} = 0$.  This suggests that our fitted function can be systematically overestimating the number of GCs from dwarf early-type galaxies with $M_V > -18$ \cite{Peng2008}.

\begin{figure}[th!]
\centering
\includegraphics[width=3.2in]{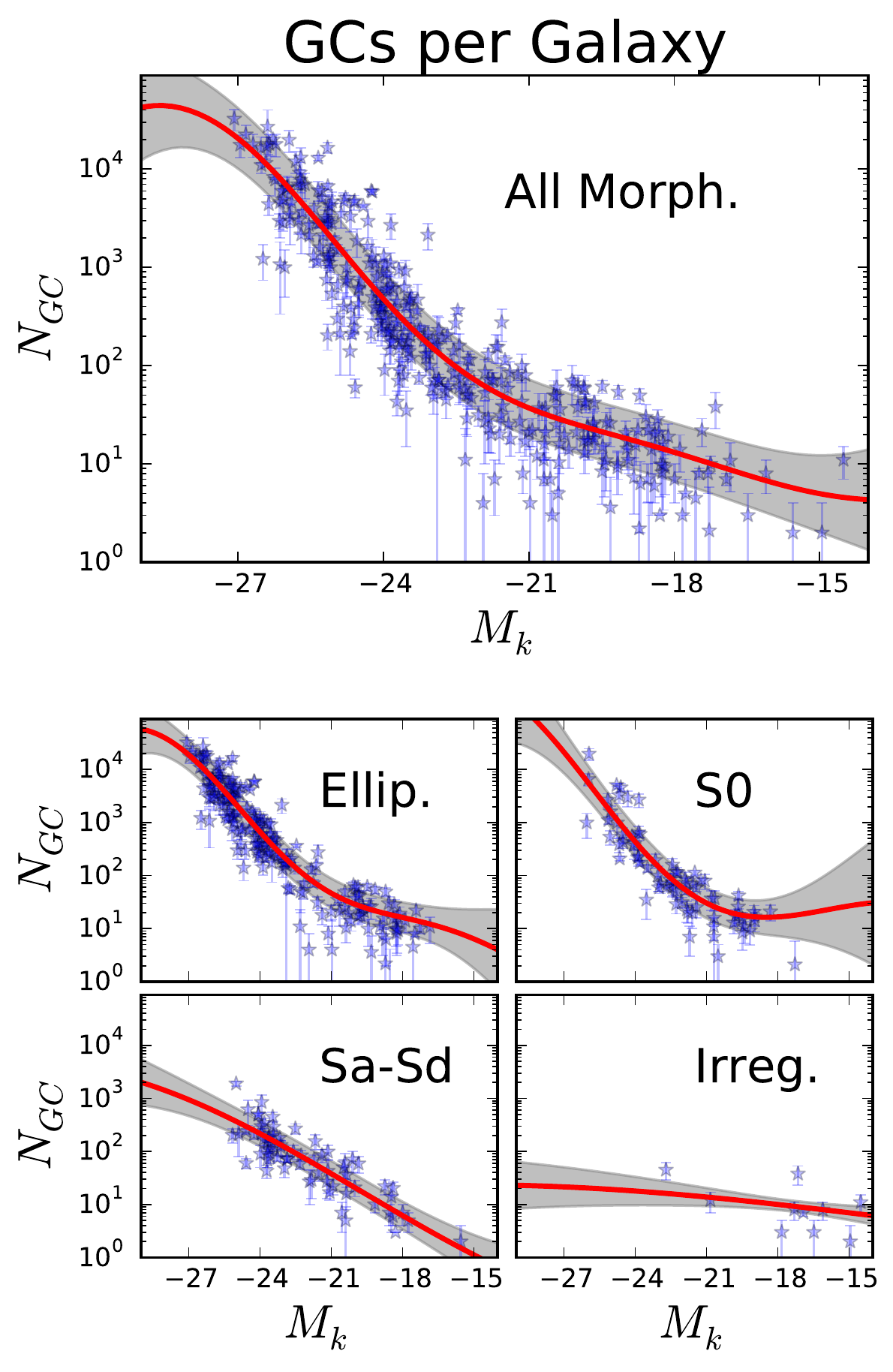}
\caption{Number of GCs per galaxy per K-band luminosity for different galaxy morphologies in the Harris GC Systems Catalog \cite{Harris2013}.  The top plot shows the fit to all galaxies in the catalog, and is the regression used for the spatial estimate of GCs in the text.  The red line shows the mean of the Gaussian process regression for each sample, while the gray shows the $1\sigma$ confidence interval about each estimate. Note that this catalog does not include observed dwarf galaxies with no GCs, suggesting that the estimated mean at low magnitudes is systematically biased to higher values.}
\label{fig:ngc}
\end{figure}

\begin{table}[b]
\begin{tabular}{l|ccc}
\hline\hline

$$ & $M_k < -17$    & $M_k < -15$    & $M_k < -13$    \\ \hline
All                 & $0.62^{1.21}_{0.32}$  & $0.77^{1.55}_{0.39}$  & $0.99^{2.31}_{0.45}$  \\ 
Elliptical          & $0.17^{0.32}_{0.09}$ & $0.17^{0.32}_{0.09}$ & $0.17^{0.33}_{0.09}$ \\ 
Lenticular          & $0.13^{0.25}_{0.07}$ & $0.13^{0.25}_{0.07}$ & $0.13^{0.25}_{0.07}$ \\ 
Spirals (Sa-Sd)     & $0.13^{0.23}_{0.08}$ & $0.13^{0.23}_{0.08}$ & $0.13^{0.23}_{0.08}$ \\ 
Irregular           & $0.11^{0.15}_{0.09}$ & $0.27^{0.35}_{0.21}$  & $0.49^{0.68}_{0.36}$   \\ \hline\hline
\end{tabular}
\caption{The number density per $\rm{Mpc}^{3}$ of GCs in the local universe, found by combining the Harris GCS catalog \cite{Harris2013} with the K-band Schechter Functions from the GAMA survey \cite{Kelvin2014}.  The errors are found by integrating the $1\sigma$ uncertainties by the Schechter functions.  Note that these errors are incomplete, as they ignore the uncertainties in the fits of the Schechter functions themselves.}
\label{tab:density}
\end{table}

The mean in Figure \ref{fig:ngc} gives us the number of GCs per galaxy per K-band luminosity.  To compute a spatial density, we must integrate the mean over the density of galaxies per K-band luminosity per $\rm{Mpc}^{-3}$.  For this, we can use the well known Schechter function \cite{Schechter1976}, which describes the spatial density of galaxies per absolute magnitude interval $dM$. We use the recently-computed K-band Schechter functions fits from the Galaxy and Mass Assembly (GAMA) Survey \cite{Kelvin2014}, which contains individual fits to individual Schechter functions for each galaxy morphology, and a single fit to a double Schechter function for the combined sample of all galaxies.  We use both to determine our overall density of GCs, as well as the density contributed by galaxies of each type.

Finally, we multiply each Schechter function from \cite{Kelvin2014} by the estimated number of GCs per galaxy from Figure \ref{fig:ngc}, and integrate over all K-band magnitudes ($M_k < -15$).  This results in a GC density estimate of $\rho_{GC} = 0.77\ \rm{Mpc}^{-3}$, which we employ in our rate calculation.  For completeness, we also consider different cutoffs for our magnitude integral, and report the contribution to the spatial densities from each galaxy morphology, in Table \ref{tab:density}.

Since the Schechter Function diverges at low luminosities, and since our fit systematically overestimates the number of GCs for low-luminosity galaxies, we must pick a reasonable limit at which to truncate our integral.  We use a lower limit of $M_k = -15$, although for completeness we also consider lower ($M_k = -17$) and higher ($M_k = -13$) cutoffs in Table \ref{tab:density}.   

In addition to comparisons with observations, we can also compute the density of GCs using cosmological simulations.  The publication of the GC Systems catalog in \cite{Harris2013} noted a correlation between the dynamical mass of a galaxy and the size of its GC population.  This relationship was expanded upon in \cite{Harris2015}, which measured a very strong correlation between the mass of the GC population and the galaxy halo mass.  This relation takes the following form:

\begin{equation}
\log_{10} M_{GCS} = \alpha + \beta(\log_{10} M_{h} - \left<\log_{10} M_{h}\right>)~,
\label{eqn:mgcs}
\end{equation}

\noindent where $\alpha$ is 7.706(7.405)(7.157), $\beta$ is 1.03(0.96)(1.21), and $\left<\log M_{h}\right>$ is 12.3(12.2)(12.2) for all(low-metallicity)(high-metallicity) GCs. Unlike the $N_{GC}-M_k$ relationship, the relationship between halo mass and $M_{GCS}$ does not strongly depend on the galaxy morphology.  

In order to convert this to a spatial density of GCs, we can multiply this relationship by the dark matter halo mass function, as determined by recent cosmological simulations.  We use the functional fit to $\frac{dn}{dM_h}$ from \cite{Tinker2010}, as calculated at redshift $z=0$ by the \texttt{HMFcalc} website \cite{Murray2013}.  We then compute the integral

\begin{equation}
\rho_{GC} = \int \frac{M_{GCS}(M_h)}{\left< M_{GC} \right>}\frac{dn}{dM_h} dM_h~,
\label{eqn:halo}
\end{equation}

\noindent where we use $\left< M_{GC} \right> = 3\times10^5M_{\odot}$, the mean of the GC mass function from \cite{Harris2014} used in the main text, to convert from the mass of a GC system to the number of GCs.  This yields a spatial density of $\rho_{GC} = 3.42~h^4\rm{Mpc}^{-3}$, or $\rho_{GC} = 0.72~\rm{Mpc}^{-3}$, assuming the value of $h = 0.677$ used throughout the text.  We can also use the similar values for low and high-metallicitiy GCs quoted below Eqn. \ref{eqn:mgcs}.  This yields estimates of $\rho_{GC}^{low} = 0.44~\rm{Mpc}^{-3}$ and $\rho_{GC}^{high} = 0.34~\rm{Mpc}^{-3}$, respectively.

\section{Models}
\label{apx:models}

This letter considers the BBH inspirals from 48 separate GC models generated with our orbit-averaged H{\'{e}}non-type Monte Carlo code, CMC.  The majority of these models were first developed in \cite{Morscher2015}.  The full details of CMC can be found in previous papers \cite{Joshi1999,Joshi2001a,Pattabiraman2013}, but the features most relevant to this letter are as follows:

\begin{itemize}
\item three-body binary formation, which we implement with a probabilistic analytic prescription \cite{Morscher2015},
\item strong single-binary and binary-binary stellar encounters, implemented with the small-$N$ integrator \texttt{Fewbody} \cite{Fregeau2004}, and
\item single and binary stellar evolution with the \text{SSE} and \text{BSE} packages \cite{Hurley2000,Hurley2002}.  Note that our implementation includes several improvements \cite{Chatterjee2010}, including the stellar remnant prescription and BH kick physics from \cite{Belczynski2002}.  For BBHs which merge within the cluster, the GW timescale is calculated by BSE.  For ejected binaries, the inspiral time is found by integrating the orbit-averaged Peters equations \cite{Peters1964} using the masses, separation, and eccentricity of the binary at the time of ejection.
\end{itemize}

The models begin with $2\times10^5$, $8\times10^5$, and $1.6\times10^6$ number of particles, and are evolved to an age of 12 Gyr each.  We do not include GC models which dissipate before 12 Gyr, as we have no way to compare these models to observations.  However, since these models all begin with low numbers of particles and produce low numbers of BBHs, the effect on the rate estimate will be minimal.  For a complete list of GC initial conditions considered, see Table \ref{tab:models}.

We also explore the space of initial cluster sizes and concentrations, with  initial virial radii of 0.5, 1, 1.5, 2, and 4 pc and initial King concentrations ($W_0$) of 2, 5, 7, and 11.  We qualitatively find that the initial King concentration does not have a strong influence on the observable GC properties at 12 Gyr.  Each of our models starts with 10\% of the objects in primordial binaries, and stellar masses chosen from a universal initial mass function (IMF) \cite{Kroupa2001}.  

We also explore metallicities of $Z=0.005$, $Z=0.001$, and $Z=0.0005$, which are placed at galactocentric distance of 2, 8 and 20 kpc respectively.  This is due to the observed correlation between GC metallicity and galactocentric distance \cite{Djorgovski1994}.  Although we explore three distinct metallicities, we separate our models into ``low metallicitiy'' (those for which [Fe/H] $\leq -0.8$, i.e. $Z=0.0005$ and $Z=0.001$), and ``high metallicity'' ($Z=0.005$) GCs.  This is chosen to simplify the comparison between our models and the observations of GCs, which show a strong bi-modality in metallicity \cite{Harris2010}.  We also assume that the fraction of low-metallicity GCs is 0.76, since that is the fraction of MWGCs for which [Fe/H] $\leq -0.8$.

\begin{table*}[h!]
\begin{tabular}{lll|llll}
\hline\hline
Initial Conditions &&& Properties (12 Gyr) \\\hline
N ($\times 10^5$) & $R_v$ (pc) & Metallicity (z) & Mass ($10^5 M_{\odot}$)& $R_c/R_h$  & BH Retained & $N_{\text{insp}}$ \\ \hline

$2$ & 0.5 & 0.001 & 0.55 & 0.17 & 0.05 & 16 \\ 
$2$ & 1.0 & 0.001 & 0.20 & 0.34 & 0.03 & 11 \\ 
$2$ & 1.0 & 0.0005 & 0.68 & 0.41 & 0.21 & 8 \\ 
$2$ & 1.5 & 0.001 & 0.54 & 0.42 & 0.11 & 8 \\ 
$2$ & 1.5 & 0.0005 & 0.62 & 0.45 & 0.18 & 7 \\ 
$2$ & 2.0 & 0.0005 & 0.63 & 0.41 & 0.20 & 5 \\ 
$2$ & 2.0 & 0.001 & 0.55 & 0.77 & 0.16 & 1 \\ 
$2$ & 2.0 & 0.0005 & 0.64 & 0.53 & 0.20 & 8 \\ 
$2$ & 2.0 & 0.001 & 0.56 & 1.12 & 0.18 & 6 \\ 
$2$ & 2.0 & 0.0005 & 0.65 & 0.45 & 0.23 & 5 \\ 
$2$ & 2.0 & 0.001 & 0.56 & 0.47 & 0.20 & 5 \\ 
$2$ & 2.0 & 0.001 & 0.50 & 0.99 & 0.22 & 3 \\ 
$2$ & 4.0 & 0.001 & 0.68 & 0.58 & 0.37 & 2 \\ 
$8$ & 0.5 & 0.0005 & 2.58 & 0.27 & 0.10 & 145 \\
$8$ & 1.0 & 0.001 & 2.17 & 0.46 & 0.18 & 91 \\ 
$8$ & 1.0 & 0.0005 & 2.78 & 0.48 & 0.24 & 87 \\
$8$ & 1.0 & 0.005 & 2.00 & 0.45 & 0.14 & 88 \\
$8$ & 1.5 & 0.0005 & 2.73 & 0.47 & 0.32 & 81 \\
$8$ & 1.5 & 0.001 & 2.61 & 0.54 & 0.30 & 67 \\
$8$ & 1.5 & 0.005 & 2.02 & 0.64 & 0.23 & 59 \\
$8$ & 2.0 & 0.0005 & 2.76 & 0.52 & 0.43 & 48 \\
$8$ & 2.0 & 0.001 & 2.62 & 0.58 & 0.40 & 53 \\ 
$8$ & 2.0 & 0.005 & 2.04 & 0.63 & 0.30 & 45 \\ 
$8$ & 2.0 & 0.0005 & 2.79 & 0.59 & 0.40 & 49 \\
$8$ & 2.0 & 0.005 & 2.00 & 0.52 & 0.32 & 45 \\ 
$8$ & 2.0 & 0.001 & 2.66 & 0.41 & 0.36 & 59 \\ 
$8$ & 2.0 & 0.0005 & 2.81 & 0.77 & 0.44 & 66 \\
$8$ & 2.0 & 0.001 & 2.70 & 0.5 & 0.40 & 57 \\ 
$8$ & 2.0 & 0.005 & 1.92 & 0.77 & 0.32 & 52 \\ 
$8$ & 2.0 & 0.001 & 2.60 & 0.67 & 0.45 & 38 \\ 
$8$ & 4.0 & 0.001 & 2.91 & 0.61 & 0.58 & 27 \\ 
$16$ & 1.0 & 0.0005 & 5.44 & 0.5 & 0.28 & 324 \\ 
$16$ & 1.0 & 0.001 & 4.69 & 0.65 & 0.27 & 269 \\
$16$ & 1.0 & 0.005 & 4.49 & 0.52 & 0.24 & 270 \\ 
$16$ & 1.5 & 0.0005 & 5.59 & 0.72 & 0.42 & 204 \\
$16$ & 1.5 & 0.001 & 5.39 & 0.71 & 0.41 & 175 \\ 
$16$ & 1.5 & 0.005 & 4.75 & 0.59 & 0.33 & 203 \\ 
$16$ & 2.0 & 0.0005 & 5.68 & 0.57 & 0.56 & 159 \\ 
$16$ & 2.0 & 0.001 & 5.50 & 0.56 & 0.49 & 154 \\ 
$16$ & 2.0 & 0.005 & 4.84 & 0.75 & 0.43 & 162 \\ 
$16$ & 2.0 & 0.0005 & 5.76 & 0.6 & 0.52 & 172 \\ 
$16$ & 2.0 & 0.001 & 5.56 & 0.66 & 0.52 & 139 \\ 
$16$ & 2.0 & 0.005 & 4.97 & 0.57 & 0.41 & 175 \\ 
$16$ & 2.0 & 0.0005 & 5.76 & 0.56 & 0.53 & 167 \\
$16$ & 2.0 & 0.001 & 5.58 & 0.67 & 0.52 & 140 \\ 
$16$ & 2.0 & 0.005 & 4.96 & 0.63 & 0.41 & 153 \\ 
$16$ & 2.0 & 0.001 & 5.47 & 0.61 & 0.52 & 121 \\ 
$16$ & 4.0 & 0.001 & 5.94 & 0.79 & 0.67 & 99 \\ \hline\hline

\end{tabular}
\caption{List of the 48 GC models used in this study.  The initial conditions are varied across the number of initial particles (N), the initial virial radius and the initial metallicities.  These models also explore a number of different initial King concentrations ($w_0$), but those are excluded from this table, as they are not observed to have a significant correlation with the observed properties at 12 Gyr.  We also include the observational properties after 12 Gyr of evolution, including the final GC mass, the ratio of the core radius to the half-light radius, the fraction of total BHs remaining in the cluster, and the total number of BBHs formed by each cluster that inspiral within 12 Gyr.}
\label{tab:models}
\end{table*}

\cleardoublepage
\clearpage


\section*{Supplementary material}

# Erratum: Binary Black Hole Mergers from Globular Clusters: Implications for Advanced LIGO [Phys. Rev. Lett. 115, 051101]

Carl L. Rodriguez,[1] Meagan Morscher,[1] Bharath Pattabiraman,[1,2]
Sourav Chatterjee,[1] Carl-Johan Haster,[1,3] and Frederic A. Rasio[1]

[1]*Center for Interdisciplinary Exploration and Research in Astrophysics (CIERA) and Dept. of Physics and Astronomy, Northwestern University, 2145 Sheridan Rd, Evanston, IL 60208, USA*
[2]*Dept. of Electrical Engineering and Computer Science, Northwestern University, Evanston, IL, USA*
[3]*School of Physics and Astronomy, University of Birmingham, Birmingham, B15 2TT, United Kingdom*
(Dated: December 21, 2015)

We have discovered an error in our integral over the globular cluster mass function that was reported in the letter. The computational error artificially over-counted the contribution from high-mass globular clusters, artificially increasing the mean number of mergers per globular cluster over 12 Gyr. This changed the quoted values in Table 1, and also increased the quoted detection rate in the paper. The corrected Table 1 is reproduced below

| | Mass Cutoff | | |
|---|---|---|---|
| **Metallicity** | $4 \times 10^6 M_\odot$ | $2 \times 10^7 M_\odot$ | $2 \times 10^8 M_\odot$ |
| Low | 159 | 223 | 250 |
| High | 238 | 424 | 549 |

TABLE I. The corrected mean number of inspirals per GC over 12 Gyr.

With the updated values from Table 1, the quoted detection rate for Advanced LIGO decreases. The realistic rate of detections is now 30 mergers per year. The pessimistic rate is 10 mergers per year. The optimistic rate is 100 mergers per year.

Fortunately, this correction significantly decreases the uncertainties quoted in our original letter. Originally, the dominant uncertainty was the upper-mass cutoff of the globular cluster mass function; however, by correctly accounting for the contribution of high-mass clusters, the upper-mass cutoff no longer heavily influences our rate estimate, and its associated uncertainty is significantly reduced. The dominant uncertainty in the updated estimate is now the spatial density of globular clusters per Mpc$^3$ in the local universe (see the first Supplemental Material). As noted in the letter, the spatial density used in previous estimates of the binary black hole merger rate is $\sim 2.4 \text{Mpc}^{-3}$ [1], similar to the number we employ in our *optimistic* rate estimate. As such, the rate we find from our models, when normalized to the same value of $\rho_{GC}$, is still $\sim 5$ times that of previous estimates [2].

This correction does not change the major conclusions of the letter, that dynamically-formed binaries contribute significantly to the overall merger rate, and that those binaries are characteristically more massive than those formed by pure stellar evolution. We reiterate that the mass difference between binaries formed dynamically and those formed in the field is heavily dependent on the physics of binary stellar evolution. In particular, we employed the original stellar wind prescription of [3], whereas more recent studies (e.g. [4]) have used the enhanced stellar wind prescription of [5]. Efforts to understand the contributions of varying binary stellar evolution physics are currently underway.

We are very grateful to Cole Miller for pointing out this error.

---



\end{document}